\font\small=cmr7
\input epsf

\def\title{
{\bf
\centerline{Chern-Simons Vortices}
}}


\def\authors{
\centerline{
C. DUVAL\foot{D\'epartement de Physique, Universit\'e
d'Aix-Marseille II and 
Centre de Physique 
Th\'eorique, \hfill\break
CNRS-Luminy, Case 907, F-13288 MARSEILLE,
Cedex 09 (France).
 e-mail:duval@cpt.univ-mrs.fr.}
and 
P. A. HORV\'ATHY
\foot{Laboratoire de Math\'ematiques et Applications,
Universit\'e de Tours, Parc de Grandmont,
F-37200 TOURS (France). e-mail: horvathy@balzac.univ-tours.fr.}
}}

\def\runningtitle{
} 

\def\runningauthors{
} 


\voffset = 1cm 
\baselineskip = 14pt 

\pageno=1
\headline ={
\ifodd\pageno\hfil\tenit\runningtitle\hfil\tenrm\folio
\else\tenrm\folio\hfil\tenit\runningauthors\hfil
\fi
} 

\nopagenumbers
\footline = {\hfil} 



\def\parag{\hfil\break} 
\def\ccr{\cr\noalign{\medskip}}
\def\IR{{\bf R}} 
\def\and{\qquad\hbox{and}\qquad}

\def\ccr{\cr\noalign{\medskip}}
\def\smallover#1/#2{\hbox{$\textstyle{#1\over#2}$}} 
\def\2{{\smallover 1/2}}
\def\L{{\cal L}}

\def\boxit#1{
\vbox{\hrule\hbox{\vrule\kern3pt
\vbox{\kern3pt#1\kern3pt}\kern3pt\vrule}\hrule}
} 
\def\dAlembert{\boxit{\null}\,} 


\newcount\ch 
\newcount\eq 
\newcount\foo 
\newcount\ref 

\def\chapter#1{
\parag\eq = 1\advance\ch by 1{\bf\the\ch.\enskip#1}
}

\def\kikezd{\parag\underbar}

\def\equation{
\leqno(\the\ch.\the\eq)\global\advance\eq by 1
}

\def\foot#1{
\footnote{($^{\the\foo}$)}{#1}\advance\foo by 1
} 

\def\reference{
\parag [\number\ref]\ \advance\ref by 1
}

\ch = 0 
\foo = 1 
\ref = 1 

\title
\vskip8mm
\authors
\vskip8mm
\noindent{\bf Abstract}.
{\it In $(2+1)$ dimensions, the Maxwell term 
$-\smallover1/4 F_{\alpha\beta}F^{\alpha\beta}$
can be replaced by the Chern-Simons three-form 
$(\kappa/4)\epsilon^{\alpha\beta\gamma}A_\alpha F_{\beta\gamma}$,
yielding a novel type of `electromagnetism'.
This has been proposed for studying the Quantum Hall Effect
as well as High-Temperature Superconductivity.
The gauge field can be coupled to a scalar field 
either relativistically or non-relativistically. 
In both cases,
one admits finite-energy, vortex solutions}.
\medskip\noindent
{\sl Contemporary Mathematics} {\bf 203}, pp. 271-288 (1997)
\bigskip

\chapter{Introduction}

The phenomenological description of `ordinary' superconductivity
is provided by Lan\-dau-\-Ginzburg theory [1]: the Cooper pairs
formed by the electrons are represented by a scalar field,
whose charge is twice that of the electron.
The scalar fields interact through their electromagnetic fields, 
governed by the
Maxwell equations. The theory admits static, vortex-like solutions
[2]. 

Landau-Ginzburg theory is non-relativistic;
a relativistic version is provided by the Abelian Higgs model [3],
which admits again static and purely magnetic vortex-type solutions.

In the eighties, however, new phenomena were discovered,
namely the 
Quantum Hall Effect [4] and High-Temperature Superconductivity [5].
None of these is described by Landau-Ginzburg theory. Hence the need
for a new theory.

Both phenomena mentioned above are essentially planar.
The mathematical room for a modification is provided by 
the three-form
$$
{\kappa\over4}\epsilon^{\alpha\beta\gamma}A_\alpha F_{\beta\gamma},
\equation
$$
introduced by Chern and Simons [6] 
in their study of secondary characteristic 
classes. This form, which only exists in $(2+1)$ dimensions, 
can be added to the conventional Maxwell term
$-\smallover1/4 F_{\alpha\beta}F^{\alpha\beta}$ [7]. 
The use of the Chern - Simons term has in fact been 
proposed in the above contexts [8], [9].

Since  the Chern - Simons term is of lower-order
in derivatives, for large distances it dominates the
Maxwell term. Thus, for describing long-range phenomena, 
the Maxwell term can even be dropped: the dynamics of the
`electromagnetic' field will be described by the
Chern-Simons term (1.1) alone.

Below, we review various aspects of Chern-Simons gauge theory.
First, we explain the construction of relativistic topological vortices.
The non-relativistic limit is
physically relevant owing to the intrinsically  non-relativistic
character of condensed matter physics. It also provides
an explicitly solvable model. Finally, we consider spinorial models. 
Again, explicit solutions are found.

\chapter{Relativistic Chern-Simons vortices}

Let us consider a complex scalar field, $\psi$, coupled to an
abelian gauge field, described by the
vector potential [1-form], $A_\mu dx^\mu$, 
defined on $(2+1)$-dimensional Minkowski space with the metric 
$(g_{\mu\nu})={\rm diag}\,(1,-1,-1)$, the coordinates being 
$x^0=t$ and $(x_i)=\vec{x}$. 
The action $S=\int d^3x\,\L$ is given by [10]
$$
\L=\2D_\mu\psi(D^\mu\psi)^*
+\smallover1/4\kappa\,\epsilon^{\alpha\beta\gamma}
A_\alpha F_{\beta\gamma}
-U(\psi),
\equation
$$
where
$
D_\mu\psi=\partial_\mu\psi-ieA_\mu\psi
$
denotes the covariant derivative, the constant $e$ is the 
electric charge of the field $\psi$, and
$\kappa$ is a  constant.
The potential $U(\psi)$ is chosen, for later convenience,
to be
$$
U(\psi)={\lambda\over4}|\psi|^2\big(|\psi|^2-1\big)^2.
\equation
$$
The Euler-Lagrange 
equations of the action (2.1) read
$$\left\{\eqalign{
&\2D_\mu D^\mu\psi=-{\delta U\over\delta\psi^*}
\equiv-{\lambda\over4}(|\psi|^2-1)\big(3|\psi|^2-1\big)\psi,
\ccr
&\2\kappa\,\epsilon^{\mu\alpha\beta}F_{\alpha\beta}=ej^\mu,
\cr}\right.
\equation
$$
where
$
F_{\alpha\beta}=\partial_\alpha A_\beta-\partial_\beta A_\alpha
$ is the `electromagnetic' field,
and
$j^\mu\equiv(\varrho,\vec{\jmath})$ is the current
$$
j^\mu={1\over2i}\big[\psi^*D^\mu\psi-\psi(D^\mu\psi)^*\big].
\equation
$$

The first of the equations in (2.3) is the Non-Linear
Klein-Gordon equation (NLKG) familiar from the Abelian Higgs model [3]; 
the second, called {\it Field-Curent Identity} (FCI), 
replaces the Maxwell equations.

It follows from the Bianchi identity,
$
\epsilon^{\alpha\beta\gamma}\,
\partial_\alpha F_{\beta\gamma}=0,
$
that the current (2.4) is conserved, 
$\partial_\mu j^\mu=0$.

Due to the presence of the vector potential in the Chern-Simons
term, the Lagrangian (2.1) 
is manifestly {\it not} invariant
with respect to a local gauge transformation
$$
\psi(x)\to e^{ie\lambda(x)}\psi(x),
\qquad
A_\mu\to A_\mu+\partial_\mu\lambda.
\equation
$$
Using the Bianchi identity, 
the change can be written, however, as
$
\partial_\alpha\left((\kappa/4)\,\epsilon^{\alpha\beta\gamma}\,
F_{\beta\gamma}\,\lambda\right).
$
This is a surface term, which does not
change the equations of motion.
The two other terms in (2.1) are gauge-invariant.

The same conclusion can also be reached by looking at
the field equations:
the FCIs are invariant, since
both the field strength and the current are invariant;
both sides of the NLKG get in turn be 
multiplied by $e^{ie\lambda}$.

\goodbreak
\kikezd{Finite-energy configurations}

Let us consider a static field configuration $(A_\mu,\psi)$.
The energy, defined as the space integral of the time-time
component of the
energy-momentum tensor associated with the Lagrangian (2.1),
is
$$
E\equiv\int\!d^2\vec{x}\,T^{00}
=\int\!d^2\vec{x}\Big[\2D_i\psi(D^i\psi)^*
-\2e^2A_0^2|\psi|^2+\kappa A_0B+U(\psi)\Big],
\equation
$$
where
$B=-F^{12}$
is the magnetic field. Note that this expression is {\it not}
positive definite.
Observe, however, that the static solutions of the equations of movement (2.3)
are  stationary points of the energy. 
Variation of (2.6) w. r. t. $A_0$ yields one of the equations of motion,
namely
$$
-e^2A_0|\psi|^2+\kappa B=0.
\equation
$$
Eliminating $A_0$ from (2.6) using this equation, 
we obtain the {\it positive definite} energy functional
$$
E=\int\!d^2\vec{r}\,\Big[\2D_i\psi(D^i\psi)^*
+{\kappa^2\over2e^2}\,{B^2\over|\psi|^2}
+U(\psi)\Big].
\equation
$$

We are interested in static, finite-energy configurations.
Finite energy at infinity is guaranteed by the conditions\foot{These 
conditions are  
in no way necessary; they yield 
the so-called {\it topological solitons}. Non-topological solutions 
are constructed in Ref. [11].}
$$
\left\{\matrix{
i.)&\qquad\vert\psi\vert^2-1
&\quad=&\quad{\rm o}\big({1/r}\big),
\ccr
ii.)
&\qquad B&\quad=&\quad{\rm o}\big({1/r}\big),
\ccr
iii.)&\qquad\vec{D}\psi
&\quad=&\quad{\rm o}\big({1/r}\big).
\cr
}\right.\qquad
{r\to\infty}.
\equation
$$
Therefore,
 the $U(1)$ gauge symmetry is broken  for large $r$.
In particular, the scalar field $\psi$ is
covariantly constant,
$\vec{D}\psi=0$.
This equation is solved  by parallel 
transport,
$$
\psi(\vec{x})=\exp\Big[i\int_{\vec{x}_0}^{\vec{x}}eA_idx^i\Big]\,\psi_0,
$$
which is well-defined whenever
$$
\oint eA_idx^i=
\int_{\IR^2}\!d^2\vec{x}\,eB
\equiv e\Phi
=2\pi n,
\qquad
n=0,\pm1,\ldots.
\equation
$$
Thus, the magnetic flux is {\it quantized}.

By i.),
the asymptotic values of the Higgs field
provide us with a mapping from
the circle at infinity
into the vacuum manifold, which is again a circle, 
$|\psi|^2=1$. 
Since the vector potential behaves asymptotically as
$$
A_j\simeq-{i\over e}\,\partial_j\log\psi,
\equation
$$
the integer $n$ in Eq. (2.10) is precisely the
{\it winding number} of this mapping; $n$ is
called the
{\it topological charge} (or {\it vortex number}).

Spontaneous symmetry breaking generates mass [12].
This occurs in a somewhat unusual guise, though.
Expanding $j^\mu$ around the vacuum expectation value of $\psi$ 
we find 
$j^\mu=-eA^\mu$ so that the FCI in (2.3) is approximately 
$$
\2\kappa\,\epsilon^{\mu\alpha\beta}F_{\alpha\beta}\simeq-e^2A^\mu.
$$
Hence
$
F_{\alpha\mu}=
-({e^2/\kappa})\,\epsilon_{\alpha\mu\beta}A^\beta.
$
Inserting here $F_{\alpha\beta}$ and deriving by $\partial^\alpha$
we find that the gauge field $A^\mu$
satisfies the Klein-Gordon equation
$$
\dAlembert A^\mu
=
-\Big({e^2\over\kappa}\Big)^2A^\mu,
\equation
$$
showing that the mass of the gauge field is
$$
m_A={e^2\over\kappa}.
\equation
$$

The Higgs mass is found in turn in the usual way: the
expectation value of the Higgs field can be chosen
as $\psi_0=(1,0)$. 
Expanding $\psi$ as $(\psi_r,\psi_\vartheta)=(1+\varphi,\theta)$
we get
$$
U=\underbrace{U(1)}_{=0}
\quad+\quad
\underbrace{
{\delta U\over\delta|\psi|}\Big\vert_{|\psi|=1}}_{=0}\,\varphi
\quad+\quad
{1\over2}\underbrace{
\left({\delta^2U\over\delta|\psi|^2}\right)
\Big\vert_{|\psi|=1}}_{m_\psi^2}
\varphi^2,
\equation
$$
since $|\psi|=1$ is a critical point of $U$. We conclude that
the mass of the Higgs particle is
$$
m_\psi^2={\delta^2U\over\delta|\psi|^2}\Big\vert_{|\psi|=1}
={2\lambda}.
\equation
$$
(This result can also be seen by considering the radial equation (2.17)
below).

\goodbreak
\kikezd{Radially symmetric solutions} [10]

For the radially symmetric Ansatz
$$
A_0=A_0(r),
\qquad
A_r=0,
\qquad
A_\vartheta=A(r),
\qquad
\psi(r)=f(r)e^{-in\vartheta},
\equation
$$
the equations of motion (2.3) read
$$
\eqalign{
&{1\over r}{dA\over dr}+{e^2\over\kappa}f^2A_0=0,
\ccr
&r{dA_0\over dr}+{e^2\over\kappa}f^2\left({n\over e}+A\right)=0,
\ccr
&{d^2f\over dr^2}+{1\over r}{df\over dr}
-{e^2\over r^2}\left({n\over e}+A\right)^2f+e^2A_0^2f
=-{\lambda\over4}f(1-f^2)(1-3f^2),
\ccr}
\equation
$$
with asymptotic conditions
$$\matrix{
\lim_{r\to\infty}A(r)\hfill&=&-\displaystyle{n\over e},\hfill
\qquad
&\lim_{r\to\infty}f(r)\hfill&=&1,\hfill
\ccr
\lim_{r\to0}A(r)\hfill&=&\ 0,\hfill
&\lim_{r\to0}f(r)\hfill&=&0.\hfill
\cr}
\equation
$$
This is either seen by a direct substitution into the equations,
or by re-writing the energy as
$$
E=2\pi\int_0^\infty dr\left\{
{r\over2}\,(f')^2
+{a^2\over 2r}f^2+{\kappa^2\over2e^4f^2}{(a')^2\over r}
+U(f)\right\},
\equation
$$
where $a=eA+n$.
The upper equation in (2.17) is plainly the radial form of (2.7).
Then variation of (2.19) with respect to $a$ and $f$ yields
the two other equations in (2.17).

Approximate solutions can be obtained by inserting the asymptotic value,
$f\simeq1$, into the first two equations: 
$$
{a'\over r}+{e^2\over\kappa}A_0=0,
\qquad
A_0'+{e^2\over\kappa}{a\over r}=0,
$$
from which we infer that
$$
{d^2A_0\over d\rho^2}
+
{1\over\rho}{dA_0\over d\rho}-A_0=0,
\qquad
\rho\equiv(e^2/\kappa)r.
$$
This is the modified Bessel equation [Bessel equation of imaginary 
argument] of order zero.
Hence
$$
A_0=CK_0({e^2\over\kappa}r).
\equation
$$
Similarly, 
for $a={n/e}+A$ we find, 
putting $\alpha=a/r$, 
$$
\alpha''+{\alpha'\over\rho}
-\Big(1+{1\over\rho^2}\Big)\alpha=0,
$$
which is Bessel's equation of order $1$ with imaginary argument.
Thus $\alpha=CK_1(\rho)$ so that
$$
A=-{n\over e}+C{e^2\over\kappa}r\,K_1\big({e^2\over\kappa}\,r\big).
\equation
$$

Another way of deriving this result is to express $A$ from
from the middle equation in (2.17),
$$
A=-{n\over e}-{\kappa\over e^2}\,r{\ d\over dr}A_0.
$$
The consistency with (2.21) follows from the recursion relation
$
K_0'=-K_1
$
of the Bessel functions. 

An even coarser approximation is
obtained by eliminating the ${a'\over r}$ 
term by setting $a=ur^{-1/2}$ and
dropping the terms with inverse powers of $r$. Then both 
equations reduce to
$$
u''=\big({e^2\over\kappa}\big)^2u
\qquad\Longrightarrow\qquad
A_0=a={C\over\sqrt{r}}\,e^{-m_Ar},
\equation
$$
which shows that the fields approach their asymptotic values
exponentially, with characteristic length determined by the
gauge field mass.

The deviation of $f$ from its asymptotic value, $\varphi=1-f$, is
found by inserting $\varphi$ into the last eqn. of (2.17);
developping to first order in $\varphi$
we get
$$
\varphi''+{1\over r}\varphi'-2\lambda\varphi\simeq0
\qquad\Longrightarrow\qquad
\varphi=CK_0(\sqrt{2\lambda}\,r),
\equation
$$
whose asymptotic behaviour is again exponential
 with characteristic length
$(m_\psi)^{-1}$,
$$
\varphi={C\over\sqrt{r}}\,e^{-m_\psi r}.
\equation
$$
The penetration depths of the 
gauge and scalar fields are therefore 
$$
\eta={1\over m_A}={e^2\over\kappa}
\qquad\hbox{and}\qquad
\xi={1\over m_\psi}={1\over\sqrt{2\lambda}},
\equation
$$
respectively.
For small $r$ instead, 
inserting the developments in powers of $r$,
we find
$$
\eqalign{
&f(r)\sim f_0r^{|n|}+\ldots,
\ccr
&A_0\sim\alpha_0-{ef_0^2n\over2\kappa|n|}\,r^{2|n|}+\ldots,
\ccr
&A\sim-{e^2f_0^2\alpha_0\over2\kappa(|n|+1)}\,r^{2|n|+2}+\ldots,
\ccr}
\equation
$$
where $\alpha_0$ and $f_0$ are constants. 
In summary, 
$$\eqalign{
|\psi(r)|&\equiv f(r)
\qquad\propto\qquad
\left\{\matrix{
r^{|n|}\hfill&\qquad r\sim0\hfill
\ccr
1-Cr^{-1/2}\,e^{-m_\psi r}
\qquad\hfill&\qquad r\to\infty\hfill\cr}\right.
\ccr
|E(r)|&=|A_0'(r)|
\quad\propto\qquad
\left\{\matrix{
r^{2|n|-1}\hfill&r\sim0\hfill
\ccr
Cr^{-1/2}\,e^{-m_Ar}+\hbox{\small lower order terms}
\qquad\hfill&r\to\infty\hfill\cr}\right.
\ccr
|B(r)|&={|A'|\over r}
\qquad\propto\qquad
\left\{\matrix{
r^{2|n|}\hfill&r\sim0\hfill
\ccr
Cr^{-3/2}\,e^{-m_Ar}+\hbox{\small lower order terms}
\qquad\hfill&r\to\infty\hfill\cr}\right.
\cr}
\equation
$$

\goodbreak
\kikezd{Self-duality}

Let us suppose that the fields have equal masses,
$m_\psi^2=m_A^2\equiv m^2$ and hence equal
penetration depths.
Then the Bogomolny trick applies, i.e., 
the energy can be rewritten in the form
$$
E=\int d^2\vec{r}\,\left[
\2|(D_1\pm iD_2)\psi|^2
+\2\Big\vert{\kappa\over e}\,{B\over\psi}
\mp{e^2\over2\kappa}\psi^*(1-|\psi|^2)\Big\vert^2\right]
\mp\int d^2\vec{r}\,{eB\over2}(1-|\psi|^2).
\equation
$$
The last term can also be presented as
$$
\mp{eB\over2}
\mp\2\vec{\nabla}\times\vec{\jmath}.
$$
The integrand of the $B$-term yields the magnetic flux; the second is
transformed, by Stokes' theorem, into the circulation of the current at
infinity which vanishes since all fields drop off at infinity
by assumption.
Its integral is therefore proportional to the magnetic flux,
$
\pm{e\Phi/2}.
$
Since the first integral is non-negative, we
have, in conclusion,
$$
E\geq {e|\Phi|\over2}=\pi|n|,
\equation
$$
equality being only attained if the `self-duality' equations
$$
D_1\psi=\mp iD_2\psi
\and
eB=\pm{m^2\over2}\,|\psi|^2(1-|\psi|^2)
\equation
$$
hold.
It is readily verified that the solutions of equations  
(2.7) and (2.31) solve
automatically the non-linear Klein-Gordon equation. 

For the radial Ansatz (2.16) the self-duality
equations become
$$
f'=\pm{a\over r}f,
\qquad
{a'\over r}=\pm\2m^2f^2(f^2-1),
\equation
$$
where we introduced again $a=eA+n$.
Deriving the first of these equations and using the 
second one, we for $f$ we get the `Liouville - type' equation 
$$
\bigtriangleup\log f={m^2\over2}\,f^2(f^2-1).
\equation
$$
 
Another way of obtaining the first-order eqns. 
(2.32) is to rewrite, for 
$$
U(f)=(m^2/8)\,f^2\big(f^2-1\big)^2,
$$
the energy as
$$
\pi\!\int_0^\infty\!rdr\left\{
\big[f'\mp{a\over r}f\big]^2
+{1\over m^2f^2}\big[{a'\over r}\mp{m^2\over2}f^2(f^2-1)\big]^2
\right\}
\pm\pi (af^2)\Big\vert_0^\infty
\mp\pi a\Big\vert_0^\infty.
$$
The boundary conditions read
$$
\matrix{a(\infty)&=0,\qquad\hfill
f(\infty)&=1,
\hfill
\ccr
a(0)&=n,\hfill
f(0)&=0,\hfill
\cr}
\equation
$$
and thus
$
E\geq\pi|n|
$
as before, with equality attained iff the equations (2.32) hold.

For $n=0$ the only solution is the vacuum,
$$
f\equiv1,
\qquad
A\equiv1.
\equation
$$ 
To see this, note that the boundary conditions at infinity are
$
f(\infty)=1
$
and
$
A(\infty)=0.
$
Let now $f(r)$, $A(r)$ denote an arbitrary finite-energy 
configuration and consider 
$$
f_\tau(r)=f(r),
\qquad
A_\tau(r)=\tau A(r)
\equation
$$
where $\tau>0$ is a real parameter. This provides us with a
$1$-parameter family configurations with finite energy
$$
E_\tau=2\pi\int_0^\infty dr\left\{
{r\over2}\,\big(f'\big)^2
+\tau^2\bigg[{a^2\over 2r}f^2
+{r\over2m^2}\big({a'\over rf}\big)^2\bigg]
+U(f)\right\},
\equation
$$
which is a monotonic function of $\tau$, whose minimum
is at $\tau=0$ i.e. for $a\equiv0$. Then Eq. (2.32) implies that
$f'\equiv0$ so that $f\equiv1$ is the only possibility.

Let us assume henceforth that $n\neq0$.
No analytic solution has been found so far. 
To study the large-$r$ behaviour, put
$\varphi\equiv1-f$.
Inserting $f\simeq1$, Eq. (2.32) reduces to 
$$
\varphi'=\mp{a\over r},
\qquad
{a'\over r}=\mp m^2\varphi.
\equation
$$
From here we get
$$\eqalign{
&\varphi''+{1\over r}\varphi'-m^2\varphi=0
\quad\Longrightarrow\quad
\varphi=CK_0(mr).
\cr
&a''-{1\over r}a'-m^2a=0
\quad\Longrightarrow\quad
a=CmrK_1(mr).
\cr}
$$
Thus, for large $r$,
$$
f\simeq1-CK_0(mr)
\and
A\simeq-{n\over e}+CmrK_1(mr)
\equation
$$
with some constant $C$. 
For small $r$ instead, Eq. (2.32), yields, to 
${\rm O}(r^{5|n|+1})$,
$$\eqalign{
&f(r)= f_0r^{|n|}
-{f_0^3m^2\over2(2n+2)^2}\,r^{3|n|+2}+{\rm O}(r^{5|n|+2}).
\ccr
&A=-{f_0^2m^2\over2(2|n|+2)e}\,r^{2|n|+2}
+{f_0^2m^2\over2(4|n|+2)e}\,r^{4n+2}+{\rm O}(r^{4|n|+4}).
\cr}
\equation
$$
The result is consistent with (2.26) since the constant 
$\alpha_0$ is now
$
\alpha_0={m/2e}={e/2\kappa}.
$

\goodbreak
\vskip5mm
\centerline{\epsfxsize=8truecm\epsfbox{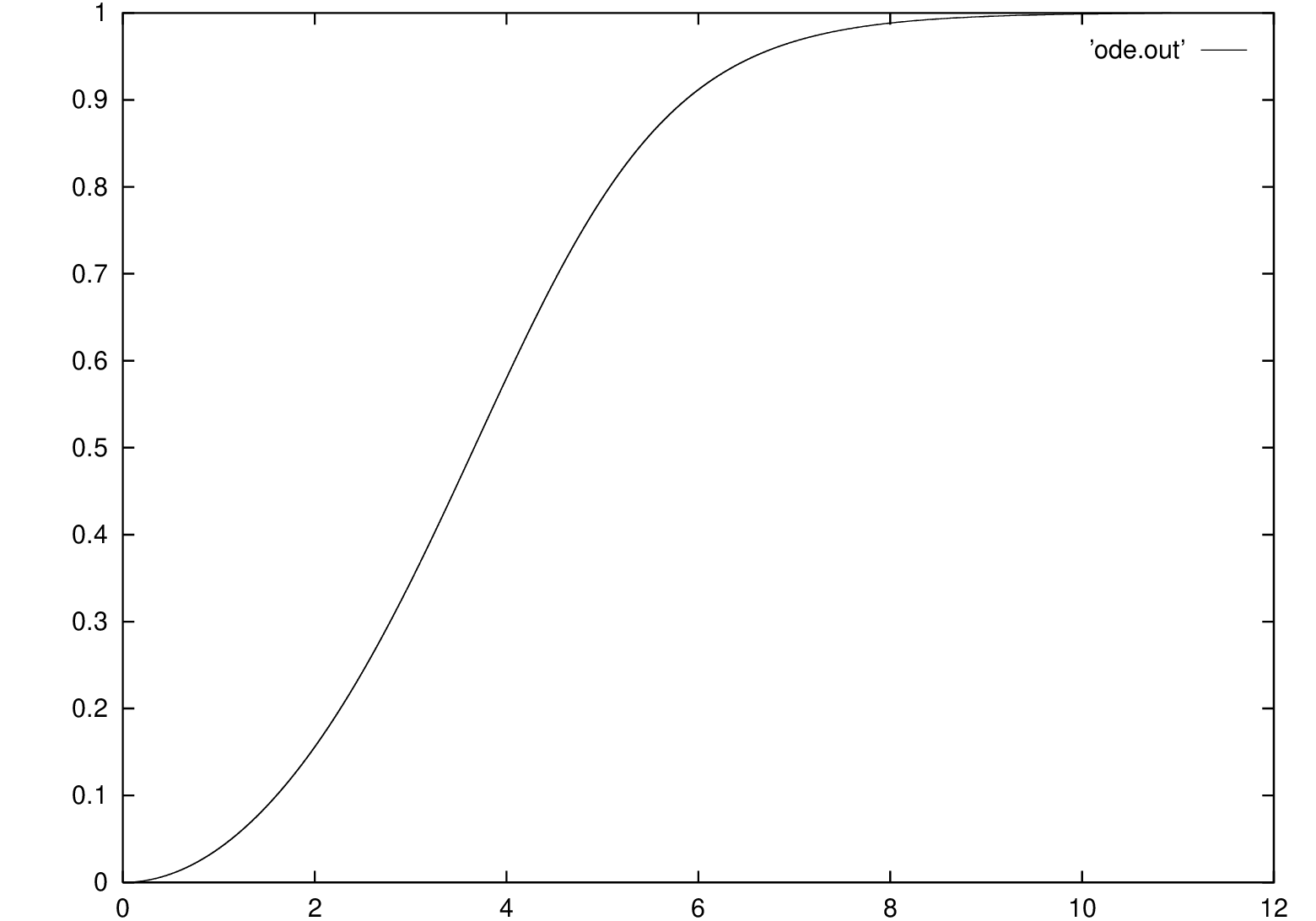}}
\vskip-0.1truecm

\kikezd{Fig. 1.}\hskip2mm
{\it The scalar field of the 
radially symmetric charge-$2$ relativistic vortex}.
\goodbreak

\null\vskip5mm
\centerline{\epsfxsize=8truecm\epsfbox{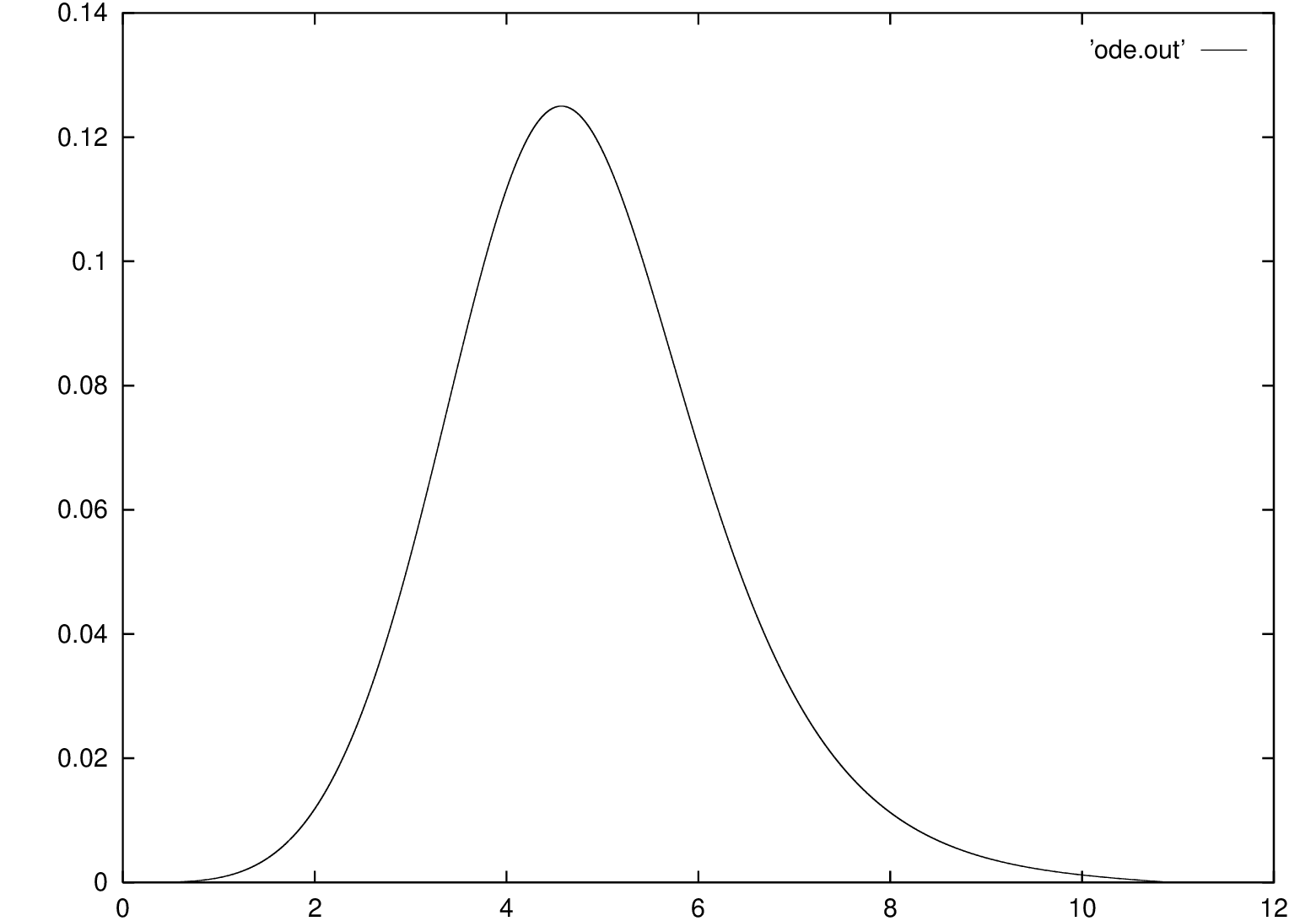}}
\noindent
\kikezd{Fig. 2.}\hskip2mm
{\it The magnetic field of the radially symmetric 
charge-$2$ relativistic vortex.
Note that $B=0$ where the scalar field vanishes: the vortex
has a doughnut-like shape}.\vskip2mm

Let us mention that 
Eq. (2.33) is actually valid in full generality, without
the assumption of radial symmetry. 
Expressing in fact the vector-potential 
from $(D_1\pm iD_2)\psi=0$ as
$$
e\vec{A}=\vec\nabla({\rm Arg}\,\psi)\pm\vec\nabla\times\log|\psi|
\equation
$$
and inserting into the second equation in (2.31), we get 
again (2.33), with $|\psi|$ replacing $f$.
Index-theoretical
calculations show that, for topological charge $n$,
Eqn. (2.33) admits a $2|n|$ parameter family of solutions [10].

\goodbreak
\chapter{Non-relativistic vortices}

The non-relativistic limit of the above system is found [13,14] by
setting
$$
\psi=e^{-imc^2t}\Psi+e^{+imc^2t}\bar\Psi,
$$
where $\Psi$ and $\bar\Psi$ denote the particles and
antiparticles, respectively. Inserting this into the
action, dropping the oscillating terms
and only keeping those of order $1/c$,
 we see that both the
particles and antiparticles are conserved. 
We can therefore consistently set $\bar\Psi=0$.
The remaining matter Lagrangian reads
$$
\L_{{\small matter}}=
i\Psi^\star D_t\Psi
-{|\vec{D}\Psi|^2\over2m}
+
{\Lambda\over2}
(\Psi^\star\Psi)^2,
\equation
$$
where $\Lambda={e^2/mc|\kappa|}$.
One can show that the theory is indeed non-relativistic [15], [16].
In what follows, we put $c=1$.

Let the constant $\Lambda$ be arbitrary.
Variation of $\int\L_{matter}$ w. r. t. $\Psi^\star$ yields the
{\it gauged non-linear Schr\"odinger equation}
$$
i\partial_t\Psi=
\left[
-{\vec{D}^2\over2m}
-eA_0
-\Lambda\Psi^\star\Psi\right]\Psi.
\equation
$$
A fully self-consistent system is obtained by
adding the matter action to the Chern-Simons action (1.1); the 
variational 
equations of this latter are clearly the  FCIs in Eq. (2.3),
written in non-relativistic notations as
$$
B\equiv\epsilon^{ij}\partial_iA^j=-{e\over\kappa}\,\varrho,
\qquad 
E^i\equiv-\partial_iA^0-\partial_tA^i=
{e\over\kappa}\epsilon^{ij}J^j,
\equation
$$
where the current is now
$$
j^\mu\equiv(\varrho,\vec\jmath)
=(\Psi^\star\Psi,{1\over2mi}\big[
\Psi^*\vec{D}\Psi-\Psi(\vec{D}\Psi)^*\big]).
\equation
$$

This is legitimate, since the 
Chern-Simons term is also {\it non-relativistic}: 
it is in fact invariant
with respect to {\it any} coordinate transformation.

\kikezd{Self-dual vortex solutions}

We would again like to find static soliton solutions.
The construction of an energy-momentum tensor is now more subtle,
because the theory is non-relativistic. This is nevertheless
possible [15], [16]. One gets the energy functional
$$
E=
\int d^2{x}\,
\Big({|\vec{D}\Psi|^2\over2m}-{\Lambda\over2}(\Psi^\star\Psi)^2\Big).
\equation
$$
Applying once more the Bogomolny trick, this is also written as
$$
E=
\int d^2{x}\,{\big|(D_1\pm iD_2)\Psi|^2\over2m}
-\2\left(\Lambda-{e^2\over m|\kappa|}\right)
(\Psi^\star\Psi)^2.
$$
Thus, for the {\it specific value} 
$$
\Lambda={e^2\over m|\kappa|}
\equation
$$
of the non-linearity,
 --- the same as the one we obtained above by taking the
non-relativistic limit of the self-dual relativistic
theory --- 
the second term vanishes. Then the energy is positive definite,
whose absolute minimum --- zero --- is attained for 
``self-dual'' fields i.e. for such that
$$
(D_1\pm iD_2)\Psi=0.
\equation
$$
Separating the phase as $\Psi=e^{ie\omega}\sqrt\varrho$, we get
$$
\vec{A}=\vec\nabla\omega
+{1\over2e}\vec\nabla\times\log\varrho,
\equation
$$
where $\varrho$ solves the Liouville equation,
$$
\bigtriangleup\log\varrho=\pm{2e^2\over\kappa}\varrho,
\equation
$$
cf. (2.33).
Physically admissible solutions arise when the r.h.s. is negative.
Hence 
the upper sign has to be chosen when $\kappa<0$ and
the lower sign when $\kappa>0$. Then the general solution
reads
$$
\varrho={4|\kappa|\over e^2}\,{|f'|^2\over[1+|f|^2]^2},
\equation
$$
where $f$ is a meromorphic function.
In particular, for 
$$
f(z)=\sum_{i=1}^n{c_i\over z-z_i},
\equation
$$
where the $c_i$ and the $z_i$ are $2n$ complex numbers,
we get a $4n$-parameter family of physically admissible
solutions, with vortex number $n$.

Note that $\varrho$ vanishes at the poles of $f$;
these points represent the \lq positions' of
the vortices. The singularity in $\varrho$ is precisely compensated by 
the singularity in the phase, leaving us with a regular magnetic
field [15].

The number of independent
parameters is in fact $4n-1$, due to a global gauge freedom;
index-theory calculations [17] show that this is
indeed the maximal number of independent solutions.

For $f=c/z^n$  in particular, we obtain the radially symmetric
solution
$$
\varrho(r)={4n^2|\kappa|\over e^2 r^2}
\left[{|c|\over r^n}+{r^n\over |c|}\right]^{-2},
\equation
$$
whose flux is {\it evenly} quantized,
$$
\Phi=-({\rm sg}\kappa){2h\over e}\times n.
\equation
$$
A few non-relativistic vortices are shown on FIGs. 3-6 below.

\vskip-15mm

\centerline{\epsfysize=9truecm\epsfbox{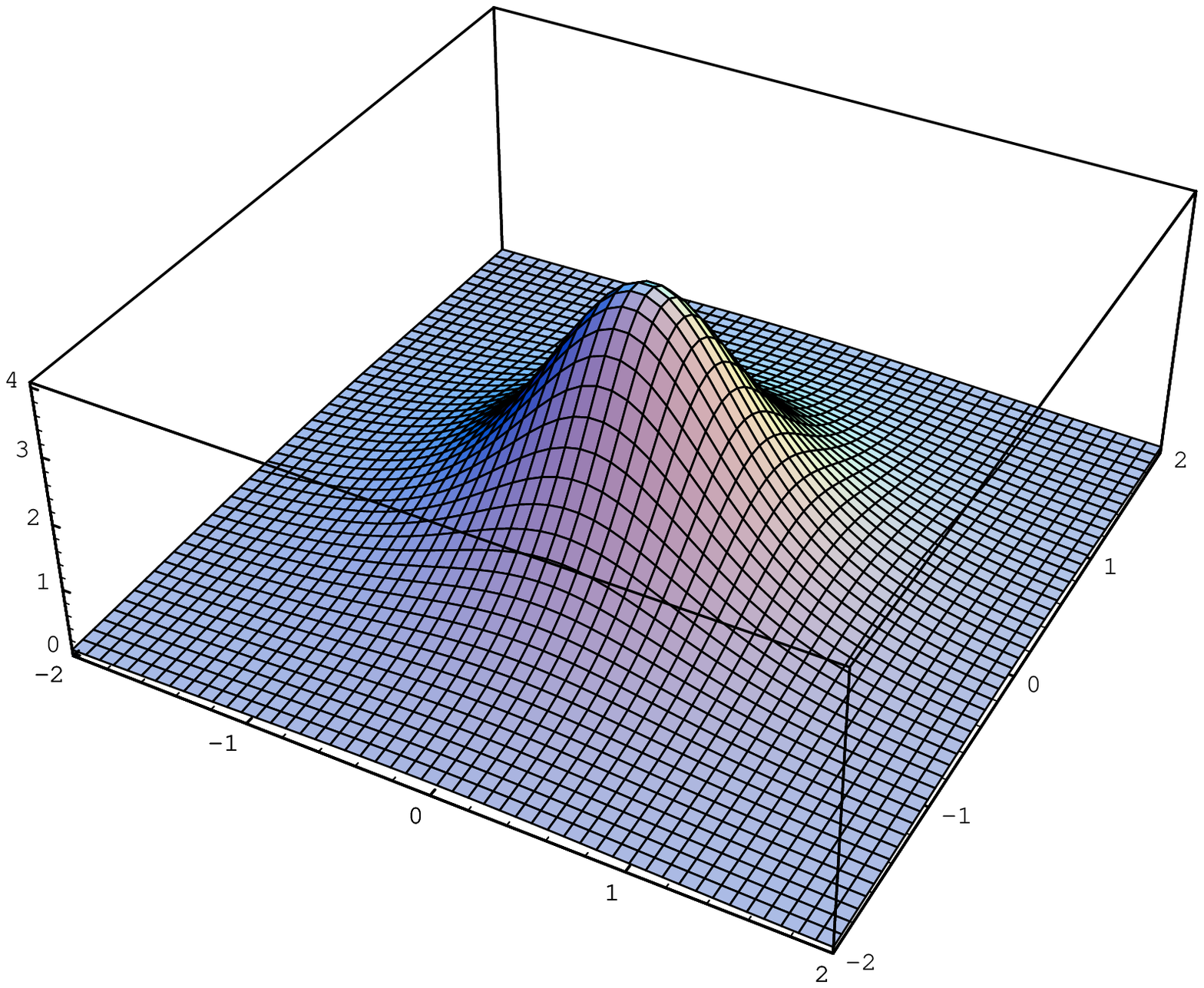}}
\vskip-25mm
\kikezd{Fig. 3}
{\it The non-relativistic radially symmetric $N=1$ vortex
has a maximum at $r=0$}.

\goodbreak
\vskip-13mm
\centerline{\epsfysize=8.5truecm\epsfbox{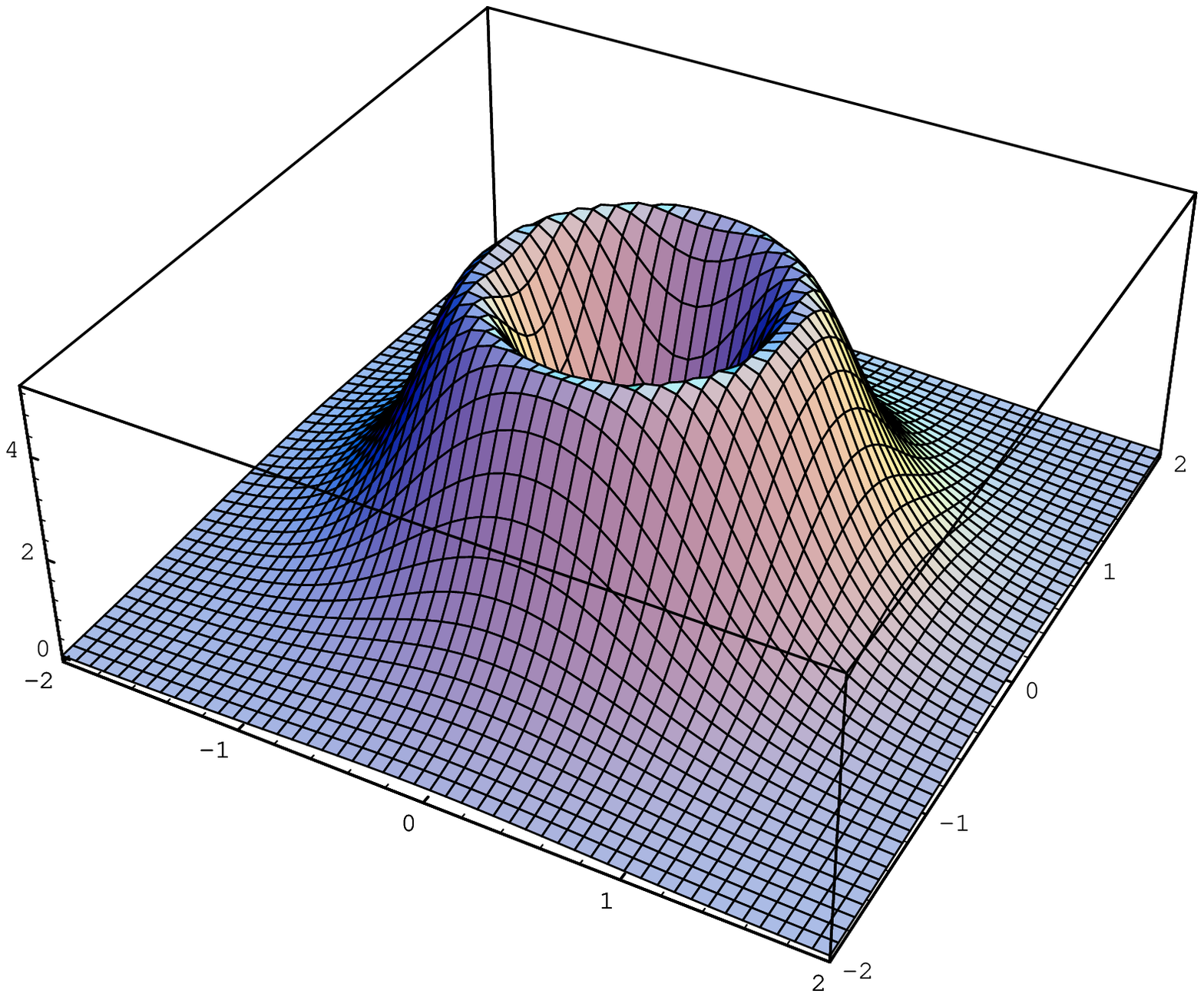}} 

\vskip-25mm
\kikezd{Fig. 4}
{\it The non-relativistic radially symmetric $N=2$ vortex
has a \lq doughnut-like' shape.
For $N\geq2$, the particle density
vanishes at $r=0$}. 

\goodbreak
\vskip-15mm

\centerline{\epsfysize=10truecm\epsfbox{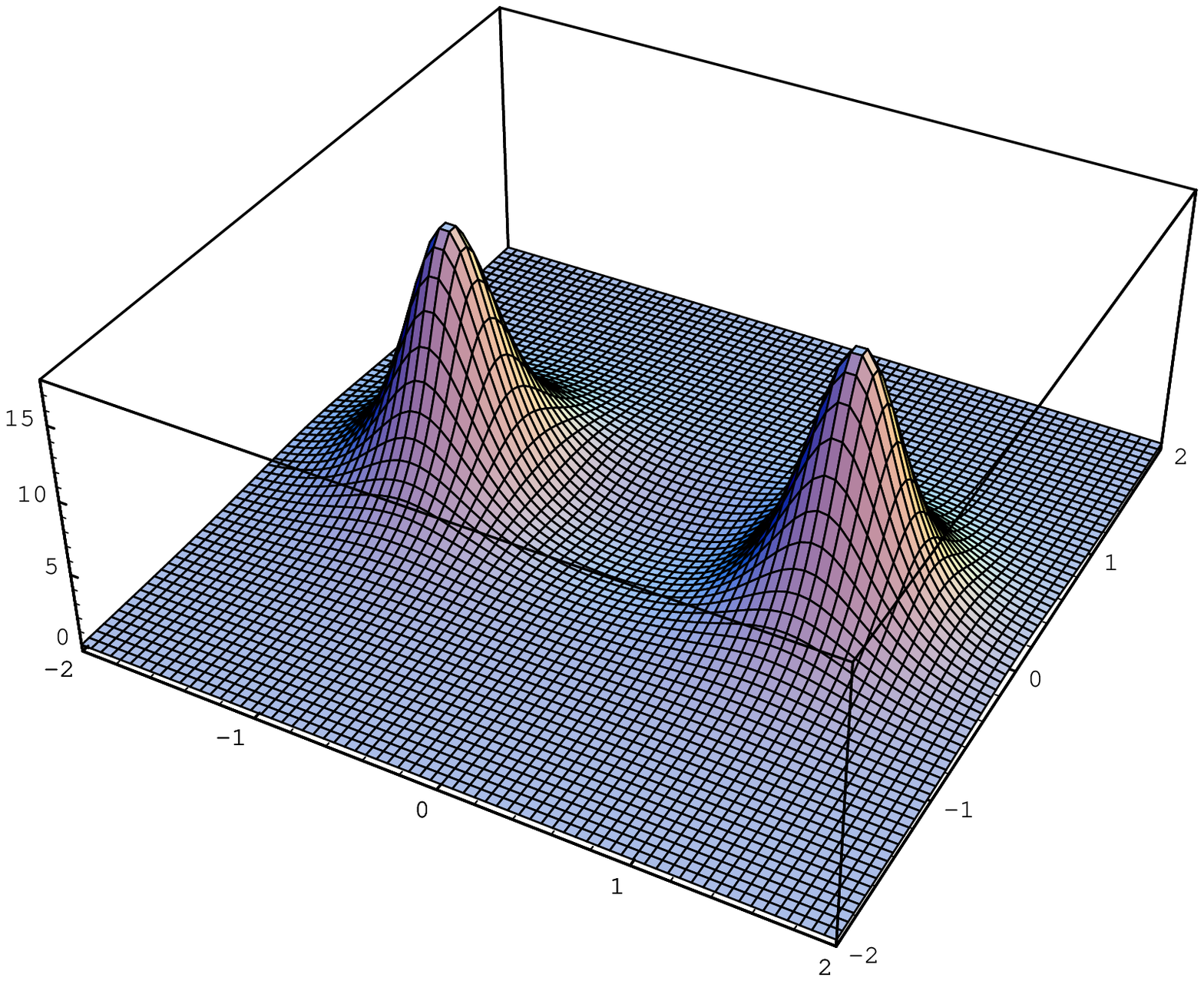}}

\vskip-25mm

\kikezd{Fig. 5}
{\it The non-relativistic $N=2$ vortex
representing two separated $1$-vortices with `positions' at $x=\pm1$ and $y=0$}.

\vskip-25mm

\centerline{\epsfysize=12.5truecm\epsfbox{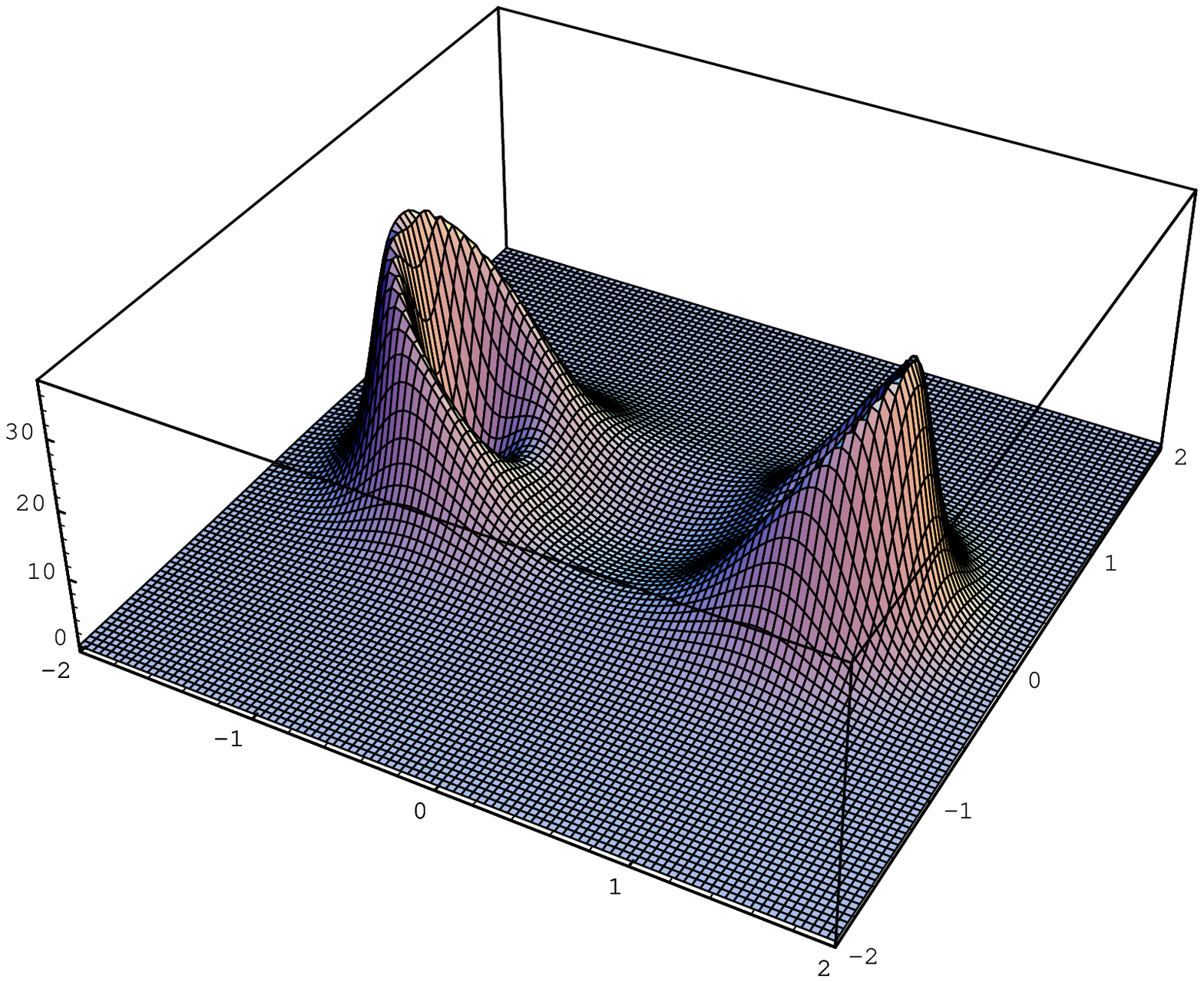}}

\vskip-29mm

\kikezd{Fig. 6}
{\it The non-relativistic $N=4$ vortex
representing two separated 
$2$-vortices with `positions' at $x=\pm1$ and $y=0$}.
\vskip2mm

Let us mention that the only time-dependent solutions found so far 
are either obtained from the static solution by the action of
the Schr\"odinger group 
(the fundamental symmetry group of the theory
[13, 14, 15, 16]), or are stationary [18].

\chapter{Spinor vortices}

\kikezd{A. Relativistic spinor vortices}

In Ref. [19] Cho et al. obtain, 
by dimensional reduction from Minkowski 
space,
a $(2+1)$-dimensional system. After some notational changes, their
equations read
$$\eqalign{
&\2\kappa\epsilon^{\alpha\beta\gamma}F_{\beta\gamma}
=
e\big(
\bar\psi_+\gamma_+^\alpha\psi_++\bar\psi_-\gamma_-^\alpha\psi_-
\big),
\ccr
&\big(ic\gamma^\alpha_\pm D_\alpha-m \big)\psi_\pm
=0,
\cr}
\equation
$$
where the two sets of Dirac matrices are
$$
(\gamma^\alpha_\pm)
=
(\pm (1/c)\sigma^3, i\sigma^2,-i\sigma^1),
\equation
$$
and the $\psi_\pm$ denote the chiral components, defined as 
eigenvectors of the chirality operator
$$
\Gamma=\pmatrix{-i\sigma_3&0\cr0&i\sigma_3\cr}.
\equation
$$
Observe that, although the Dirac equations are decoupled, the
chiral components are nevertheless coupled through the Chern-Simons 
equation.
Stationary solutions, representing purely magnetic vortices, 
are readily found [19].
It is particularly interesting to construct {\it static} solutions.
For $A_0=0$ and $\partial_tA_i=0$, setting
$$
\psi_{\pm}=e^{-imt}\pmatrix{F_\pm\cr G_\pm\cr},
\equation
$$
the {\it relativistic} system (4.1) becomes, for $c=1$,
$$\eqalign{
&\kappa\epsilon^{ij}\partial_iA_j=-e\big(|F_+|^2+|G_-|^2\big),
\ccr
&\big(D_1+iD_2\big)F_\pm=0,
\qquad
\big(D_1-iD_2\big)G_\pm=0.
\cr}
\equation
$$
Now, for $F_\pm=0$ or $G_\pm=0$, these equations are
{\it identical} to those which describe the 
{\it non-relativistic, self-dual} vortices of Jackiw and Pi [15]~!

\kikezd{B. Non-relativistic spinor vortices}

Non-relativistic spinor vortices
can also be constructed along the same lines [20].
Following L\'evy-Leblond [21], 
a non-relativistic spin $\2$ field $\psi=\pmatrix{\Phi\cr\chi}$ 
where  $\Phi$ and $\chi$ are two-component `Pauli' spinors,
is described by the $2+1$ 
dimensional equations
$$\left\{
\matrix{
(\vec{\sigma}\cdot\vec{D})\,\Phi
&+\hfill&2m\,\chi&=&0,
\ccr
D_t\,\Phi&+\hfill&i(\vec{\sigma}\cdot\vec{D})\,\chi\hfill&=&0.
\cr}\right.
\equation
$$

These 
spinors are coupled to the Chern-Simons gauge field through 
the mass (or particle) density,
$
\varrho=|\Phi|^2,
$ 
as well as through the spatial components of the current, 
$$
\vec{J}=i\big(\Phi^\dagger\vec{\sigma}\,\chi
-\chi^\dagger\vec{\sigma}\,\Phi\big),
\equation
$$
according to the Chern-Simons FCI (2.3).
The chirality operator is still given by Eq. (4.3)
and is still conserved. Observe that
$\Phi$ and $\chi$ in Eq. (4.6) are {\it not} the chiral components 
of $\psi$; these latter are defined by 
$
\2(1\pm i\Gamma)\psi_\pm=\pm\psi_\pm.
$

It is easy to see that Eq. (4.6) 
splits into two uncoupled systems for  
$\psi_+$ and $\psi_-$.
Each of the chiral components separately describe (in 
general different) physical phenomena in $2+1$ dimensions. 
For the ease of presentation, 
we keep, nevertheless, all four components of $\psi$.

Now the current can be written in the form:
$$
\vec{J}={1\over2im}\Big(
\Phi^\dagger\vec{D}\Phi-(\vec{D}\Phi)^\dagger\Phi\Big)
+\vec{\nabla}\times\Big({1\over2m}\,\Phi^\dagger\sigma_3\Phi\Big).
\equation
$$
Using 
$(\vec{D}\cdot\vec{\sigma})^2=
\vec{D}^2+eB\sigma_3$,
we find that the component-spinors satisfy
$$\left\{
\matrix{
&iD_t\Phi&=&-{1\over2m}\Big[\vec{D}^2+eB\sigma_3\Big]\Phi,\hfill
\ccr
&iD_t\chi&=&-{1\over2m}\Big[\vec{D}^2+eB\sigma_3\Big]\chi
-\smallover{e}/{2m}\,(\vec{\sigma}\cdot\vec{E})\,\Phi.
\cr}\right.
\equation
$$
Thus, $\Phi$ solves a \lq Pauli equation', while   
$\chi$ couples through the Darwin term,
$\vec{\sigma}\cdot\vec{E}$. 
Expressing $\vec{E}$ and $B$ through the FCI, (2.3) 
and inserting into our equations, 
we get finally 
$$\left\{
\matrix{
&iD_t\Phi=
&\Big[-{1\over2m}\,\vec{D}^2
+{e^2\over2m\kappa}\,|\Phi|^2\,\sigma_3
\Big]\Phi,\hfill
\ccr
&iD_t\chi=
&\Big[-{1\over2m}\,\vec{D}^2
+\smallover e^2/{2m\kappa}\,|\Phi|^2\,\sigma_3
\Big]\chi
-\smallover e^2/{2m\kappa}\,\big(
\vec{\sigma}\times\vec{J}\big)\Phi.
\cr}\right.
\equation
$$
If the chirality of $\psi$ is restricted to $+1$ (or $-1$),
this system describes 
non-relativistic spin $+\2$ ($-\2$) fields interacting with a 
Chern-Simons gauge field. 
Leaving the chirality of $\psi$ unspecified, it 
describes {\it two} spinor fields of spin $\pm\,\2$, 
interacting with each other
and the Chern-Simons gauge field.

Since the lower component is simply
$
\chi=-(1/2m)(\vec{\sigma}\cdot\vec{D})\Phi,
$
it is enough to solve the
$\Phi$-equation.
For 
$$
\Phi_+=\pmatrix{\Psi_+\cr0\hfill\cr}
\and
\Phi_-=\pmatrix{0\hfill\cr\Psi_-\cr}
\equation
$$
respectively --- which amounts to working with the $\pm$ chirality
components --- the \lq Pauli' equation in (4.10)
reduces to
$$
iD_t\Psi_\pm=
\Big[-{\vec{D}^2\over2m}
\pm\lambda\,(\Psi_\pm^\dagger\Psi_\pm)\Big]\Psi_\pm,
\qquad
\lambda\equiv{e^2\over2m\kappa},
\equation
$$
which again (3.2), but with non-linearity
$\pm\lambda$, {\it half} of the 
special value $\Lambda$ in (3.6) used by Jackiw and Pi.
For this reason, our solutions (presented below) 
will be {\it purely magnetic}, ($A_t\equiv0$), unlike in the case
studied by Jackiw and Pi. 
In detail,
let us consider the static system
$$\left\{\eqalign{
&\Big[-{1\over2m}(\vec{D}^2+eB\sigma_3)-eA_t\Big]\Phi=0,
\ccr
&\vec{J}=-{\kappa\over e}\vec\nabla\times A_t,
\ccr
&\kappa B=-e\varrho,
\cr}\right.
\equation
$$
and try the first-order Ansatz
$$
\big(D_1\pm iD_2\big)\Phi=0.
\equation
$$ 
Eq. (4.14) makes it possible to replace $\vec{D}^2=D_1^2+D_2^2$
by $\mp eB$, then the first
equation in (4.13) can be written as
$$
\Big[-{1\over2m}eB(\mp 1+\sigma_3)-eA_t\Big]\Phi=0,
\equation
$$
while the current is 
$$
\vec{J}
=
{1\over2m}\vec\nabla\times\Big[\Phi^\dagger(\mp1+\sigma_3)\Phi\Big].
\equation
$$

Now, due to the presence of $\sigma_3$, both
Eq. (4.16) and the second equation in (4.13) can be solved
with a {\it zero} $A_t$ and $\vec{J}$: 
by choosing 
$\Phi\equiv\Phi_+$ 
($\Phi\equiv\Phi_-$) for the upper (lower) cases respectively
makes $(\mp 1+\sigma_3)\Phi$ vanish. 
(It is readily seen from Eq. (4.15) that any solution has
a definite chirality).

The remaining task is to solve the first-order conditions
$$
(D_1+iD_2)\Psi_+=0,\qquad{\rm or}\qquad(D_1-iD_2)\Psi_-=0,
\equation
$$
which is done in the same way as in Section 3:
$$\vec{A}=\pm{1\over2e}\vec{\nabla}\times\log\varrho,
\qquad
\bigtriangleup\log\varrho=\pm{2e^2\over\kappa}\varrho.  
\equation
$$
A normalizable solution is obtained for
$\Psi_+$ when $\kappa<0$, and for $\Psi_-$ when $\kappa>0$.
(These correspond precisely to having an attractive non-linearity
in Eq. (4.12)). 
The lower components  vanish in both cases, as seen from
$$
\chi=-{1\over 2m}(\vec{\sigma}\cdot\vec{D})\Phi.
\equation
$$ 
Both solutions only involve 
{\it one} of the $2+1$ dimensional spinor fields $\psi_\pm$, depending 
on the sign of $\kappa$.

The physical properties such as symmetries and conserved quantities 
can be studied by noting that our 
equations are in fact obtained by variation of
the $2+1$-dimensional action given in [20],
which can also be used to 
show that the coupled L\'evy-Leblond --- Chern-Simons system is,
just like its scalar counterpart, Schr\"odinger symmetric [15].

A conserved energy-momentum tensor can be constructed and used to
derive conserved quantities [20]. One finds that
the `particle number' $N$ determines the actual values of all 
the conserved charges: for the radially symmetric solution, e.g.,
the magnetic flux, $-eN/\kappa$, 
and the mass, ${\cal M}=mN$,  
are the same as for the scalar soliton of [15]. 
The total angular 
momentum, however, can be shown to be $I=\mp N/2$,
{\it half} of the corresponding value 
for the scalar soliton. 
As a consequence of self-duality,
our solutions have {\it vanishing energy}, just like the ones of 
Ref. [15].

\kikezd{C. The non-relativistic limit}

Setting
$$
\psi_+=e^{-imc^2t}\pmatrix{\Psi_+\cr\widetilde{\chi}_+\cr}
\and
\psi_-=e^{-imc^2t}\pmatrix{\widetilde{\chi}_-\cr\Psi_-\cr},
\equation
$$
Eq. (4.1) become
$$\left\{\eqalign{
&iD_t\Phi-c\vec{\sigma}\cdot\vec{D}\widetilde{\chi}=0,
\ccr
&iD_t\widetilde{\chi}+c\vec{\sigma}\cdot\vec{D}\Phi
+2mc^2\widetilde{\chi}
=0,
\cr}\right.
\equation
$$
where
$
\Phi=\pmatrix{\Psi_+\cr\Psi_-\cr}
$
and
$\widetilde{\chi}=\pmatrix{\widetilde{\chi}_-\cr\widetilde{\chi}_+\cr}.
$
In the non-relativistic limit  
$mc^2\widetilde{\chi}>>iD_t\widetilde{\chi}$,
so that this latter can be dropped from
the second equation. Redefining $\widetilde{\chi}$ as
$\chi=c\widetilde{\chi}$ yields precisely our
Eq. (4.6).
This also explains, why one gets the same (namely the Liouville)
equation both in 
the relativistic and the non-relativistic cases:
for static and purely magnetic fields, the terms containing $D_t$ 
are automatically zero. 

In this paper, we only reviewed the Abelian Chern-Simons
theories. Non-Abelian generalizations are studied in
Refs. [14], [22]  and [23].
\goodbreak

\kikezd{Acknowledgement}.
We are indebted to our collaborators
S. Nicolis (Tours) and
L. Palla (Budapest),
as well as to R. Jackiw (MIT) for many interesting discussions.


\goodbreak
\vfill\eject
\centerline{\bf References}

\reference 
See, e.g., E. M. Lifshitz and L. PL. Pitaevski,
{\sl  Statistical Physics}, Landau-Lifshitz Course of Theoretical Physics,
Vol. 9. Part 2. Butterworth-Heinemann (1995).

\reference 
A. A. Abrikosov, {\sl Sov. Phys. JETP} {\bf 5}, 1174 (1957).

\reference
H. B. Nielsen and P. Olesen, 
{\sl Nucl. Phys}. {\bf B61}, 45 (1973).

\reference 
K. v. Klitzing, G. Dorda, and M. Pepper,
{\sl Phys. Rev. Lett}. {\bf 45}, 497 (1980);
D. C. Tsui, H. L. Stormer, and A. C. Gossard,
{\sl Phys. Rev. Lett}. {\bf 48}, 1559 (1982).

\reference 
J. G. Bednorz and K. A. M\"uller,
{\sl Z. Phys}. {\bf B64}, 189 (1986).

\reference 
S. S. Chern, {\sl  Complex Manifolds Without Potential Theory},
2nd ed., Springer, N. Y. (1979).

\reference
R. Jackiw and S. Templeton, 
{\sl Phys. Rev}. {\bf D23}, 2291 (1981);
J. Schonfeld, 
{\sl Nucl. Phys}. {\bf B185}, 157 (1981);
S. Deser, R. Jackiw and S. Templeton, 
{\sl Phys. Rev. Lett}. {\bf 48}, 975 (1982); 
{\sl Ann. Phys}. {\bf 140}, 372 (1982).

\reference 
S. M. Girvin, {\sl  The quantum Hall Effect}, 
ed. R. E. Prange and S. M. Girvin. Chap. 10.
(Springer Verlag, New York (1986);
S. C. Zhang, T. H. Hansson and S. Kivelson,
{\sl Phys. Rev. Lett}. {\bf 62}, 82 (1989);
J. Fr\"ohlich and U. Studer,
{\sl Rev. Mod. Phys}. {\bf 65}, 733 (1993).

\reference 
Y.-H. Chen, F. Wilczek, E. Witten and B. I. Halperin, 
{\sl Int. Journ. Mod. Phys}. {\bf B3} 1001 (1989).

\reference
S. K. Paul and A. Khare,
{\sl Phys. Lett}. {\bf B174}, 420 (1986);
J. Hong, Y. Kim and P. Y. Pac,
{\sl Phys. Rev. Lett}. {\bf 64}, 2230 (1990);
R. Jackiw and E. Weinberg, 
{\sl Phys. Rev. Lett}. {\bf 64}, 2234 (1990).

\reference
R. Jackiw, K. Lee and E. Weinberg, 
{\sl Phys. Rev}. {\bf D42}, 2234 (1990).

\reference
S. Deser and Z. Yang,
{\sl Mod. Phys. Lett}. {\bf A4}, 2123 (1989).

\reference
R.~Jackiw and S-Y.~Pi,
{\sl Phys. Rev}. {\bf D42}, 3500 (1990);
{\sl Prog. Theor. Phys. Suppl}. {\bf 107}, 1 (1992).

\reference
G. Dunne,
{\sl Self-dual Chern-Simons solitons},
Springer Lecture Notes in Physics (New Series)
(1995).

\reference
R.~Jackiw and S-Y.~Pi,
{\sl Phys. Rev. Lett}. {\bf 64}, 2969 (1990).

\reference
C. Duval, P. A. Horv\'athy and L. Palla, 
{\sl Phys. Lett}. {\bf B325}, 39 (1994).

\reference
S. K. Kim, K. S. Soh, and J. H. Yee,
{\sl Phys. Rev}. {\bf D42}, 4139 (1990).

\reference
K. H. Cho, D-H. Oh, C. Rim,
{\sl Phys. Rev}. {\bf D46}, 2709 (1992).

\reference 
Y. M. Cho, J. W. Kim, and D. H. Park,
{\sl Phys. Rev}. {\bf D45}, 3802 (1992).

\reference
C. Duval, P. A. Horv\'athy and L. Palla,
{\sl Phys. Rev}. {\bf D52}, 4700 (1995);
 {\sl Ann. Phys}. (N. Y.). {\bf 249} 265 (1996).

\reference
J-M. L\'evy-Leblond, 
{\sl Comm. Math. Phys}. {\bf 6}, 286 (1967).

\reference
H. J. De Vega and F. A. Schaposnik,
{\sl Phys. Rev. Lett}. {\bf 56}, 2564 (1986);
B. Grossmann, {\sl Phys. Rev. Lett}. {\bf 65}, 3230 (1990);
G. V. Dunne, R. Jackiw, S-Y Pi, and C. A. Trugenberger,
{\sl Phys. Rev}. {\bf D43}, 1332 (1991).

\reference
L. Martina, O. K. Pashaev, and G. Soliani,
{\sl Phys. Rev}. {\bf B48}, 15 787 (1993).

\bye